\documentclass[12pt]{article}
\usepackage{amssymb}

\usepackage{amsmath}

\begin{document}

\begin{center}
GENERALIZED HAMILTON FUNCTION IN THE PHASE SPACE OF COORDINATES AND THEIR
MULTIPLE DERIVATIVES

\textrm{Timur F. Kamalov}

\textit{Physics Department, Moscow State Open University}

\textit{P. Korchagina 22, Moscow 107996, Russia}

\textit{E-mail: ykamalov@rambler.ru}

\textit{qubit@mail.ru}
\end{center}

\textit{Refined are the known descriptions of particle behavior with the
help of Hamilton function in the phase space of coordinates and their
multiple derivatives. This entails existing of circumstances when at closer
distances gravitational effects can prove considerably more strong than in
case of this situation being calculated with the help of Hamilton function
in the phase space of coordinates and their first derivatives. For example,
this may be the case if the gravitational potential is described as a power
series in 1/r. At short distances the space metrics fluctuations may also be
described by a divergent power series; henceforth, these fluctuations at
smaller distances also constitute a power series, i.e. they are functions of
1/r. For such functions, the average of the coordinate equals zero if the
frame of reference coincides with the point of origin.}

\begin{center}

1.Introduction.
\end{center}

We can suppose that the kinematic formula

\begin{center}
$s=s_{0}+v\Delta t+\frac{a\Delta t^{2}}{2}$
\end{center}

is follow by Taylor's decomposition

\begin{center}
$s=s_{0}+v\Delta t+\frac{a\Delta t^{2}}{2}+\frac{1}{3!}\overset{\cdot }{a}%
\Delta t^{3}+\frac{1}{4!}\overset{\cdot \cdot }{a}\Delta t^{4}+...$
\end{center}

if the acceleration equal to constant.

Newtonian Physics is used Lagrangian $L=L(q,p)$. Here we consider
Non-Newtonian case when Lagrangian is depend of coordinates and their
multiple derivatives $L=L(q,\overset{\cdot }{q},\overset{\cdot \cdot }{q},%
\overset{\cdot (3)}{q},...,\overset{\cdot (n)}{q})$. The trajectory $r=r(t)$
of the classical particle can be representative by Taylor's decomposition in 
$t_{0}$

\begin{center}
$r(t-t_{0})=r(t_{0})+\overset{\cdot }{r(t_{0})}(t-t_{0})+\frac{1}{2}\overset{%
\cdot \cdot }{r(t_{0})}(t-t_{0})^{2}+\frac{1}{3!}\overset{\cdot \cdot \cdot }%
{r(t_{0})}(t-t_{0})^{3}+...$
\end{center}

Than

\begin{center}
$r(t-t_{0})=\sum_{n=0}^{N}\frac{1}{n!}\overset{\cdot (n)}{r(t_{0})}%
(t-t_{0})^{n}$.
\end{center}

The first tree quantitates from Taylor's decomposition is ordinary used in
kinematic. If to consider the other quantitative than it is very difficult
to find the all solutions. Because form Newton's pioneer's physical works it
is limited three quantitates only. On that time in general cases there are
specific movement of the particles where \ impossible to use $\overset{\cdot
\cdot \cdot }{r}=\overset{\cdot (4)}{r}=...=\overset{\cdot (n)}{r}=0$.

Lets accept the denote of the generalized derivation

\begin{center}
$\overset{\cdot (n)}{\nabla }=(\frac{\partial }{\partial \overset{\cdot (n)}{%
r}}),$

$\overset{\cdot (n)}{\nabla _{\beta }}=(\frac{\partial }{\partial \overset{%
\cdot (n)}{r^{\beta }}}).$
\end{center}

The Generalized Momentum $p$ is

\begin{center}
$p_{\alpha }=\sum_{n=0}^{N}(-1)^{n-\alpha }\frac{d^{n-\alpha }}{dt^{n-\alpha
}}(\frac{\partial L}{\partial r_{\alpha }^{\cdot (n)}})=%
\sum_{n=0}^{N}(-1)^{n-\alpha }\frac{d^{n-\alpha }}{dt^{n-\alpha }}\overset{%
\cdot (n)}{\nabla _{\alpha }}L$,
\end{center}

where the Lagrangian is the function of coordinates and their multiple
derivatives $L=L(r,\overset{\cdot }{r},\overset{\cdot \cdot }{r},\overset{%
\cdot \cdot \cdot }{r},...\overset{\cdot (n)}{r})$

The Generalized Energy for the general case is the function of multiple
derivatives

\begin{center}
$E=E_{1}+E_{2}+E_{3}+..=$

$=(ar^{2}+b\overset{\cdot }{r}^{2})+(c\overset{\cdot \cdot }{r}^{2}+d\overset%
{\cdot \cdot \cdot }{r}^{2})+(g\overset{\cdot (4)2}{r}+h\overset{\cdot (5)2}{%
r})+....$
\end{center}

here $a,b,c,d,...$ is coefficients and $E_{1},E_{2},E_{3},...$ is energy of
first rang, second rang, third rang and so on.

The first quantitative is named the potential energy. The second - the
kinetic energy. The others is possible name as the energy of the second
derivation, the energy of the third derivation and so on.

Than the Generalized Lagrangian's equation is

\begin{center}
$\sum_{n=0}^{N}(-1)^{n+1}\frac{d^{n}}{dt^{n}}(\frac{\partial L}{\partial 
\overset{\cdot (n)}{r}})=\sum_{n=0}^{N}\frac{d^{n}}{dt^{n}}\overset{\cdot (n)%
}{\nabla }L=0$
\end{center}

or

\begin{center}
$\frac{\partial L}{dr}-\frac{d}{dt}(\frac{\partial L}{\partial \overset{%
\cdot }{r}})+\frac{d^{2}}{dt^{2}}(\frac{\partial L}{\partial \overset{\cdot
\cdot }{r}})-\frac{d^{3}}{dt^{3}}(\frac{\partial L}{\partial \overset{\cdot
\cdot \cdot }{r}})+\frac{d^{4}}{dt^{4}}(\frac{\partial L}{\partial \overset{%
\cdot (4)}{r}})+...=0$\bigskip
\end{center}

or in our case

\begin{center}
$\sum_{n=0}^{N}(-1)^{n+1}\frac{d^{n}}{dt^{n}}\frac{dL}{d\overset{\cdot (n)}{r%
}}=0$
\end{center}

because

\begin{center}
$\delta S=\delta \int L(r,\overset{\cdot }{r},\overset{\cdot \cdot }{r},%
\overset{\cdot \cdot \cdot }{r},...r^{(\cdot n)})dt=$

$=\int \sum_{n=0}^{N}(-1)^{n}\frac{d^{n}}{dt^{n}}(\frac{\partial L}{\partial 
\overset{\cdot (n)}{r^{\beta }}})\delta r^{\beta }dt=\int ((-1)^{n}\frac{%
d^{n}}{dt^{n}}\overset{\cdot (n)}{\nabla _{\beta }}L)\delta r^{\beta }dt=0$
\end{center}

From the Generalized Lagrangian's equation it is follow the Generalized
Newtonian's equation

\begin{center}
$F+F^{(1)}+F^{(2)}+...=\frac{dp}{dt}+\frac{d^{2}p^{(1)}}{dt^{2}}+\frac{%
d^{3}p^{(2)}}{dt^{3}}+...$
\end{center}

or

\begin{center}
$\sum_{\alpha =0}^{\infty }F^{(\alpha )}=\sum_{\alpha =0}^{\infty }\frac{%
d^{\alpha +1}p^{(\alpha )}}{dt^{\alpha +1}}$
\end{center}

if we denote

\begin{center}
$F=\frac{dp}{dt},$ where $F=\frac{\partial L}{\partial r},p=\frac{\partial L%
}{\partial \overset{\cdot }{r}}$

$F^{(1)}=\frac{dp^{(1)}}{dt},$ where $F^{(1)}=\frac{\partial L}{\partial 
\overset{\cdot \cdot }{r}},p^{(1)}=\frac{\partial L}{\partial \overset{\cdot
\cdot \cdot }{r}}$

$F^{(2)}=\frac{dp^{(2)}}{dt},$ where $F^{(2)}=\frac{\partial L}{\partial 
\overset{\cdot (4)}{r}},p^{(2)}=\frac{\partial L}{\partial \overset{\cdot (5)%
}{r}}$

..................

$F^{(\alpha )}=\frac{dp^{(\alpha )}}{dt},$ where $F^{(\alpha )}=\frac{%
\partial L}{\partial \overset{\cdot (2\alpha )}{r}},p^{(\alpha )}=\frac{%
\partial L}{\partial \overset{\cdot (2\alpha +1)}{r}}$.
\end{center}

The Generalized action function $S$ with coefficients $p,q,l,m,...$\ is

\begin{center}
$S=\sum_{\alpha =0}^{\infty }p_{\alpha }\overset{\cdot (\alpha )}{r}%
=\sum_{\alpha =0}^{\infty }\sum_{n=0}^{N}(-1)^{n}\overset{\cdot (\alpha )}{r}%
\frac{d^{n}}{dt^{n}}(\frac{\partial L}{\partial \overset{\cdot (n)}{%
r^{\alpha }}})=$

$=\sum_{\alpha =0}^{\infty }\sum_{n=0}^{N}\overset{\cdot (\alpha )}{r}%
(-1)^{n}\frac{d^{n}}{dt^{n}}\overset{\cdot (n)}{\nabla _{\alpha }}L$
\end{center}

or

\begin{center}
$S=L_{1}t+L_{2}t^{2}+...=(-pr^{2}+q\overset{\cdot }{r}^{2})t+(-l\overset{%
\cdot \cdot }{r}^{2}+m\overset{\cdot \cdot \cdot }{r}^{2})t^{2}+...$
\end{center}

here $L_{1},L_{2},L_{3},...$ is energy of first rang, second rang, third
rang and so on.

Usually, the Hamilton function[1] is expressed as a function in the phase
space of coordinates and their first derivatives. However, it can be assumed
that there exist such complex types of particle motion when these are to be
described by multiple coordinate derivatives. Such motion types can
comprise, for instance, fluctuations particle motion. There exists also an
Appel[2] definition introducing a so-called acceleration energy, which is a
quadratic form in coordinate first derivatives. Therefore, the Hamilton
function can be written as a sum of the quadratic form of coordinates,
expressing the system potential energy, the quadratic form of first
coordinate derivatives, expressing the kinetic energy, and the quadratic
form of second coordinate derivatives, which is the Appel acceleration
energy. The Generalized Hamilton function shall take on the form

\begin{center}
$H=\sum_{\alpha =0}^{\infty }p_{\alpha }\overset{\cdot (\alpha )}{r}%
=\sum_{\alpha =0}^{\infty }\sum_{n=0}^{N}(-1)^{n}\overset{\cdot (\alpha )}{r}%
\frac{d^{n}}{dt^{n}}(\frac{\partial L}{\partial \overset{\cdot (n)}{%
r^{\alpha }}})=$

$=\sum_{\alpha =0}^{\infty }\sum_{n=0}^{N}(-1)^{n}\overset{\cdot (\alpha )}{r%
}\frac{d^{n}}{dt^{n}}\overset{\cdot (n)}{\nabla _{\alpha }}L$
\end{center}

We shall call this function the generalized Hamilton function.

This function shall be studied in the phase space of coordinates and their
multiple derivatives.

\begin{center}
\bigskip\ 2. Scalar potential in the phase space of coordinates and their
multiple derivatives.
\end{center}

The Generalized Poisson's equation for the scalar potential $\varphi ^{\cdot
(n)}$ of gravitational field in this case from the sources with density
distribution of the source $\rho $ and factor $\varkappa $ depending on the
system of units shall take on the form

\begin{center}
$\sum_{n=0}^{N}(-1)^{n}\frac{d^{n}}{dt^{n}}(\frac{\partial \varphi }{%
\partial \overset{\cdot (n)}{r}})=\varkappa \rho $
\end{center}

or, in our case, Generalized Poisson's equation is

\begin{center}
$\sum_{n=0}^{N}(-1)^{n}\frac{d^{n}}{dt^{n}}\overset{\cdot (n)}{\nabla }%
\varphi =\varkappa \rho $.
\end{center}

Than the solution of Generalized Poisson's equation is

\begin{center}
$\varphi =\varphi _{0}\exp k/r$,
\end{center}

named the Generalized Green's function.

Its solution is

\begin{center}
$\varphi =\varphi ^{1}(\frac{k}{r-r_{0}})+\varphi ^{2}(\frac{k^{2}}{%
(r-r_{0})^{2}})+\varphi ^{3}(\frac{k^{3}}{(r-r_{0})^{3}})+...+\varphi ^{N}(%
\frac{k^{N+1}}{(r-r_{0})^{N+1}})$.
\end{center}

For the gravitational field the potential is

\begin{center}
$\varphi =\frac{GM}{k}\exp k/r$,
\end{center}

where $\varphi $- potential, $G$- gravitational constant, $k$- unknown
constant, which \textit{may be equal to the gravitational radius} $r_{g}$, $%
M=\int \rho dv$ - mass, $r=x-x_{0}<<1$, $x$ and $x_{0}$- coordinates. The
constant $k$ is unknown, but if $k$ is equal to the Plank constant $%
l_{p}=10^{-33}$ cm than this potential is always the same as Newtonian
potential $\varphi =GM/r$. If constant $k$ is equal to the size of nuclear $%
k=10^{-15}$ m than the gravitational force is equal to nuclei forces because
at the this distant gravitational forces is change on exponential law and
may be stronger than electromagnetic forces. The electromagnetic forces it
is possible to express by exponential characters of the interaction too. But
constants will be another in this case.

\begin{center}
3. Conclusion
\end{center}

From this paper follow, that the phase space of coordinates and there
multiple derivative gives the corrected Newton's formula for gravitational
potential $\varphi $ of two mass $m$\ is

\begin{center}
$\varphi =Gm(a\frac{k}{r}+b\frac{k^{2}}{r^{2}}+c\frac{k^{3}}{r^{3}}+...)$,
\end{center}

where $a$,$b$,$c$,... - constants.

Here $k$ is the unknown constant which have the seance of distance. For
example, if $k\thicksim 10^{-15}m$ and more\ we have always Newtonian low.

For long distances $r>>k$, we have the equation for the gravitational
potential $\varphi =Gm\frac{1}{r}$, where $k=\frac{1}{a}$.

For particle described by the generalized Hamilton function at small
distances, i.e. when the series diverges, there shall be much stronger
forces acting than it is usually considered in calculations employing the
Hamilton function. This theory of short-range interaction explains
interaction of bodies at small distances and refines the description of
their interaction in case of their increase. It can be supposed that this
method can be applied to cases when the force of gravitational attraction of
particles described by the Hamilton function, at low distances.

In this approach it is clear that the exponential characters of  
gravitational and elecromagnetic interaction completed the description of 
dark energy.

\begin{center}
References
\end{center}

1. V.I. Arnold (2000), Math. Methods of Class. Mechanic's, Editorial
URCC(Moscow).

2. P. Appel (1960), Theor. Mechanic's, Fizmatgiz(Moscow).

\end{document}